\documentclass[a4paper,12pt]{article}
\usepackage{epsfig}
\begin{document}
\begin{titlepage}

\begin{flushright}
       LYCEN 2002-63 \\
       November 15th, 2002 \\
\end{flushright}

\vfill

{\LARGE\bf \begin{center}
SICANE: a Detector Array for the Measurement
        of Nuclear Recoil Quenching Factors
        using a Monoenergetic Neutron Beam
\end{center}}
\vfill

\begin{center}
 E.~Simon$^{1}$,
 L.~Berg\'e$^{2}$,
 A.~Broniatowski$^{2}$,
 R.~Bouvier$^{1}$,
 B.~Chambon$^{1}$,
 M.~De~J\'esus$^{1}$,
 D.~Drain$^{1}$,
 L.~Dumoulin$^{2}$,
 J.~Gascon$^{1}$,
 J-P.~Hadjout$^{1}$,
 A.~Juillard$^{2}$,
 O.~Martineau$^{1}$,
 C.~Pastor$^{1}$,
 M.~Stern$^{1}$ and
 L.~Vagneron$^{1}$
\end{center}

{\noindent\small
$^{1}$IPN Lyon, IN2P3-CNRS and Universit\'e Claude Bernard Lyon I,
      4 rue Enrico Fermi, F-69622 Villeurbanne, Cedex, France \\
$^{2}$CSNSM, IN2P3-CNRS, Universit\'e Paris XI,
      B\^at. 108, F-91405 Orsay Cedex, France
}
\vfill

\begin{center}{\large\bf Abstract}\end{center}

SICANE is a neutron scattering multidetector facility for the
determination of the quenching factor (ratio of the response to
nuclear recoils and to electrons) of cryogenic detectors used
in direct WIMP searches.
Well collimated monoenergetic neutron beams are
obtained with inverse $(p,n)$ reactions.
The facility is described, and results obtained for the quenching
factors of scintillation in NaI(Tl) and of heat and ionization in
Ge are presented.

\vfill


\end{titlepage}


\section{Introduction}

Since 1985~\cite{bib-witten},
various experimental techniques have been proposed to detect the
nuclear recoils that would be produced by the scattering of WIMP
dark matter particles from our galactic halo (for a
recent review of experimental searches,
see e.g. Ref.~\cite{bib-expreview}).
The goal is to be sensitive to recoil energies down to
10 keV~\cite{bib-lewin},
at event rates well below one per kilogram of detector
material per day~\cite{bib-rate}.
For this, the detector response to low-energy nuclear recoils
must be well known.
Depending on the detection process
(ionization, scintillation or heat),
this response may differ significantly from that inferred from
calibration with electron or gamma-ray sources.
A quantity of interest is thus the {\em quenching factor},
defined as the ratio of the signal amplitudes induced by
a nuclear recoil and an electron of the same energy.
This factor depends mainly on the detector
material, the energy of the recoil and the detection process,
although temperature and alignment effects may play a role
(see e.g. Ref.~\cite{bib-align}).

Quenching factor measurements have been made on
scintillation detectors such as
NaI~\cite{bib-nai},
CsI~\cite{bib-csi},
CaF$_2$~\cite{bib-caf2}
and Xe~\cite{bib-xe},
leading to values ranging from 4 to 30\%
depending on the recoil nucleus and its energy.
The measured quenching factors of ionization detectors such as
Si~\cite{bib-si} or
Ge~\cite{bib-chasman,bib-jones,bib-shutt,bib-messous}
vary between 25 and 40\%
and appear to follow the recoil energy dependence
predicted by the Lindhart theory~\cite{bib-lindh}.
More recently, the thermal detection efficiency for recoiling nuclei has
been measured for the first time in a TeO$_{2}$ cryogenic detector,
giving a thermal quenching factor slightly above unity\cite{bib-aless}.

The purpose of the SICANE facility (SIte de CAlibration NEutron)
is the measurement of quenching factors of
cryogenic detectors using nuclear recoils
induced by monoenergetic neutron beams.
The facility has been commissioned with
tests using scintillation in a NaI(Tl) crystal
and measurements of quenching factors
for ionization and heat signals
in a germanium cryogenic detector
developed by the EDELWEISS collaboration.

\section{Experimental Procedure}

\subsection{Principle and Motivation}

The basic method for the measurement of the quenching factor
of a detector is the following.
First the detector response to electron recoils is
obtained using calibrated $\gamma$-ray source.
This yields the calibration of the signal amplitude
in terms of an {\em electron-equivalent\/} energy, $E_{ee}$.
In a second step, the detector is exposed to a source
producing nuclear recoils with a known kinetic energy $E_R$.
The quenching factor $Q$ is then
\begin{equation}
Q = \frac{E_{ee}}{E_R}
\end{equation}
The main difficulty for a precise and efficient measurement
is the production of recoils with known $E_R$ values.

In the present method, nuclear recoils are produced by neutron
scattering. Other methods for producing nuclear recoils exist,
such as the implantation of $\alpha$-particle sources close to the
surface of the detector~\cite{bib-aless,bib-zhou}. When the $\alpha$
escapes the detector, the response of the detector to the
recoiling daughter can be measured accurately.
However this method is not adequate for detectors which response
differs according to whether the interaction occurs close to surface
or in the bulk of the volume~\cite{bib-benoit}.
Another drawback is that the recoiling daughter is often very
different from the detector material.
Neutron interactions have the advantage of occurring
rather uniformly throughout the detector volume.
The recoil nuclei thus produced are very similar to those that
would be induced by WIMP scattering.

In Ge cryogenic detectors with simultaneous read-out
of the heat and ionization signals the
recoil energy can be measured event-by-event,
independently of the nature of the recoil
(electron or nuclear).
In this type of detector, it is possible
to directly measure the {\em ratio\/} of the
ionization and heat quenching
values\footnote{For clarity, from now on the symbol
for ionization or scintillation quenching factors
will be $Q$, and $Q'$ will be used to represent
heat quenching factors.}
as a function of recoil energy~\cite{bib-shutt}.
If one assumes that the heat quenching factor $Q'$ is
unity\footnote{This assumption is reasonable if
the detector can be approximated as a close system
where all the excitation energy is thermalized
within a time scale inferior to that of the
read-out of the heat signal and no energy is stored
in crystal defects or other long-life processes.},
these data can be interpreted as ionization quenching
measurements.
Alternatively, these data can be compared to the available
direct measurements of ionization quenching in order to provide
a verification of how close to unity are the heat
quenching factors~\cite{bib-shutt}.
This comparison assumes that the ionization quenching
factors do not depend on temperature, as all direct
measurements~\cite{bib-chasman,bib-jones,bib-messous}
were performed at
liquid nitrogen temperature (77 K) instead of
cryogenic temperatures ($\sim$25 mK).

This method cannot be applied to detectors with a
single read-out, 
or to heat-and-scintillation detectors,
as the amount of energy escaping as light
is difficult to control.
In addition, scintillation yields are known to depend
on crystal temperature (see e.g. Ref.~\cite{bib-temp}).

Clearly, a method providing nuclear recoils of
a known energy 
is relevant for the calibration of cryogenic detectors.
In a non-relativistic elastic neutron scattering, the energy
of the recoil nuclei $E_R$ is given by
\begin{equation}
E_{R}(\theta) = E_{n} \frac{4Mm_n}{(m_n+M)^{2}}
                      \frac{(1-\cos\theta)}{2}
\label{eqn-ern}
\end{equation}
where $E_n$ is the incident neutron kinetic energy,
$\theta$ is the neutron scattering angle in the center-of-mass
frame and $m_n$ and $M$ are the neutron and nucleus masses,
respectively.
Even with a monoenergetic neutron beam, $E_R$ values
vary from zero to $E_R(180^o)$.
In experiments where $\theta$ is not measured some information
about the quenching factor can be obtained
from the maximum $E_{ee}$ value observed for a
given $E_n$, the quenching factor being then equal to
$E_{ee}/E_R(180^o)$~\cite{bib-chasman}.
A more precise measurement consists in measuring the scattering
angle of the neutron~\cite{bib-chasman}.
This requires neutron detectors with a small solid angle.
However, in order to be efficient the total solid angle for the
detection of the scattered neutrons must be as large as possible.
This is a crucial point for performing in a reasonable time
quenching factor measurements of cryogenic heat detectors,
as their response is intrinsically slow.

These precision and efficiency constraints naturally lead to
a concept of a large array of neutron detectors.
With 48 neutron detectors, SICANE is the largest existing array.
With such an extensive array, scattering at different
angles can be measured simultaneously, reducing the systematic
errors, and improving the diagnostic tools to
identify the various background sources.

The basic performance of this array can readily be tested
using scattering on a small NaI detector at room temperature,
for which scintillation quenching data already
exist~\cite{bib-nai}.
For actual measurements on cryogenic detectors,
two new potential problems arise.
Firstly, the cryostat surrounding the detector
represents an important amount of additional material
where neutrons may scatter before or after
interacting in the detector itself.
Secondly, cryogenic detectors are intrinsically slow,
and some care must be taken in establishing
clean coincidences with
the relatively faster array of neutron detectors
while retaining a sufficiently high count rate.
For the commissioning measurements of an actual cryogenic
detector, a Ge heat-and-ionization detector was chosen
as the experimental situation
for these detectors is rather well controlled (see above),
while still awaiting direct measurements of
the low temperature behavior of both $Q$ and $Q'$.

\subsection{Neutron Beams}

Up to now, experiments involving these techniques used
monoenergetic neutrons produced by deuteron beam interactions on a
deuterated target, or by a proton beam interacting on a $^{7}$Li
target~\cite{bib-messous}.
The neutron emission in these reactions is almost isotropic.
This reduces the flux impinging on the target detector,
and imposes that the scattered-neutron detector array be very
carefully shielded from direct neutrons~\cite{bib-putte},
putting severe constraints on the spatial extension of the array.

In contrast, reactions with inverse kinematics (heavy ion beam
on hydrogenated target) produce a naturally collimated neutron
emission~\cite{bib-dave,bib-drosg}.
Using endothermal reaction enhances the focusing of the
neutron emission in the forward direction.
The neutron flux on the detector under study is thus
increased while the flux on the surrounding array is reduced.
The neutrons are all produced within a small opening angle
$\Theta_{max}$ given by:
\begin{equation}
\sin^{2}\Theta_{max} = \frac{m_p M_f}{m_n M_i}(1-\frac{E_{th}}{E_i})
\end{equation}
where
E$_{th}$ is the threshold energy,
E$_i$ is the incident kinetic energy, and
$M_i$, $m_p$, $m_n$ and $M_f$ are the masses of the projectile,
the proton, the neutron and the residual nucleus, respectively.
In terms of the (negative) $Q$-value of the reaction,
$E_{th}$ is equal to $-Q(M_i+m_p)/m_p$.
At a given angle relative to the beam axis $\Theta<\Theta_{max}$,
there are two monoenergetic neutron groups with energies
$E_{n1}$ and $E_{n2}$ given by:
\begin{eqnarray}
E_{n1,n2} & = & -\frac{2 Q M_i}{x (m_p+M_i)}
\{
x + \cos^2\Theta \pm \sqrt{\cos^2\Theta (x + \cos^2\Theta)}
\}
\label{eqn-en}
\end{eqnarray}
where $x$ = $E_{th}$ / $E_i$,
and terms of order $Q/M_i$ and $(m_n-m_p)/M_i$
have been neglected.

The optimization of the neutron beam has been performed in
experiments conducted at the tandem accelerator facilities of the
Institut de Physique Nucl\'eaire d'Orsay.
Beams of $^{7}$Li~\cite{bib-dave} and $^{11}$B~\cite{bib-drosg}
have been tested using an
oscillating hydrogenated polyethylene target with a
thickness of 200 $\mu$g$\cdot$cm$^{-2}$.
The beams were pulsed at 2.5 MHz with 5~ns wide bursts.

\begin{figure}[tbp]
\epsfig{file=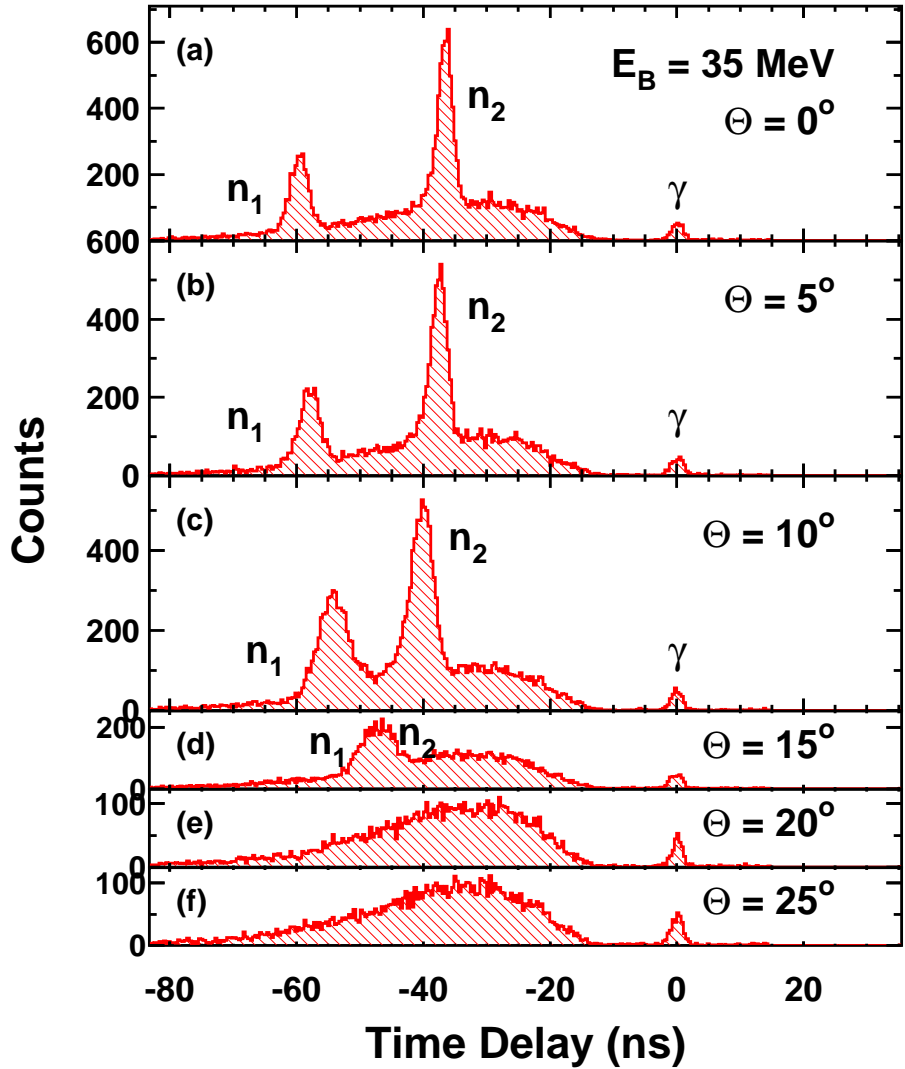
,height=17cm,bbllx=4cm,bblly=5cm,bburx=24cm,bbury=25cm}
\caption[]{
     Time of flight spectra between the arrival of the $^{11}$B beam
     burst on the hydrogenated target and the detection of a particle
     in a NE213 neutron detector located at $\Theta$ angles varying
     from 0$^o$ to 25$^o$. The incident $^{11}$B beam energy is 35 MeV.}
\label{fig-btofang}
\end{figure}

\begin{figure}[tbp]
\epsfig{file=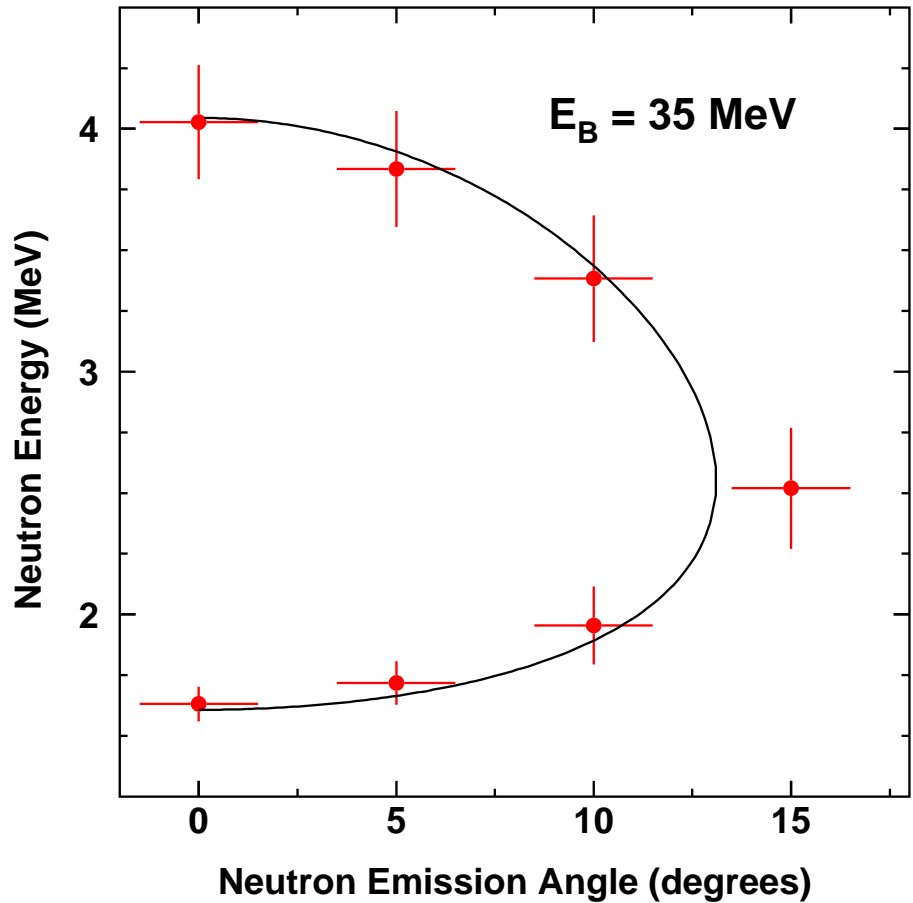,width=14cm}
\caption[]{
        Neutron energy as a function of the emission angle
        for the reaction p($^{11}$B,$^{11}$C)n
        at 35 MeV beam energy.
        The line is the expectation from Eq.~\protect\ref{eqn-en}.
        The experimental points corresponds to the peaks
        observed in the preceding figure.
        The horizontal error bar correspond to the solid angle of
        the neutron detector and the vertical one to the
        neutron energy spread associated to the width
        of the observed peaks.}
\label{fig-bang}
\end{figure}

The contents of the induced neutron beams were studied by measuring
the time-of-flight between the beam bursts and the arrival in a
NE213 scintillation detector
located at an angle of $\Theta$ varying from 0 to 25
degrees at a distance of 110~cm away from the hydrogenated target.
The time-of-flight spectra recorded at a $^{11}$B beam energy of
35 MeV for different values of $\Theta$ are shown in
Fig.~\ref{fig-btofang}.
The two neutron groups (labeled $n_1$ and $n_2$)
and the prompt $\gamma$ produced in reactions in the
hydrogenated target (labeled $\gamma$) are clearly seen.
The positions of the $n_1$ and $n_2$ peaks follow the expectation of
Eq.~\ref{eqn-en}, as illustrated in Fig.~\ref{fig-bang}, merging
together when the maximum value $\Theta_{max} = 13.1^o$ is
reached. In addition, an important background under the $n_1$ and
$n_2$ peaks is observed in Fig.~\ref{fig-btofang}.
This continuum remains
even beyond the $\Theta_{max}$ value imposed by the kinematics of
the inverse (p,n) reaction. It is also present at all beam
energies. It is therefore attributed to neutrons induced in fusion
reactions on $^{12}$C nuclei from the polyethylene target.

\begin{figure}[tbp]
\epsfig{file=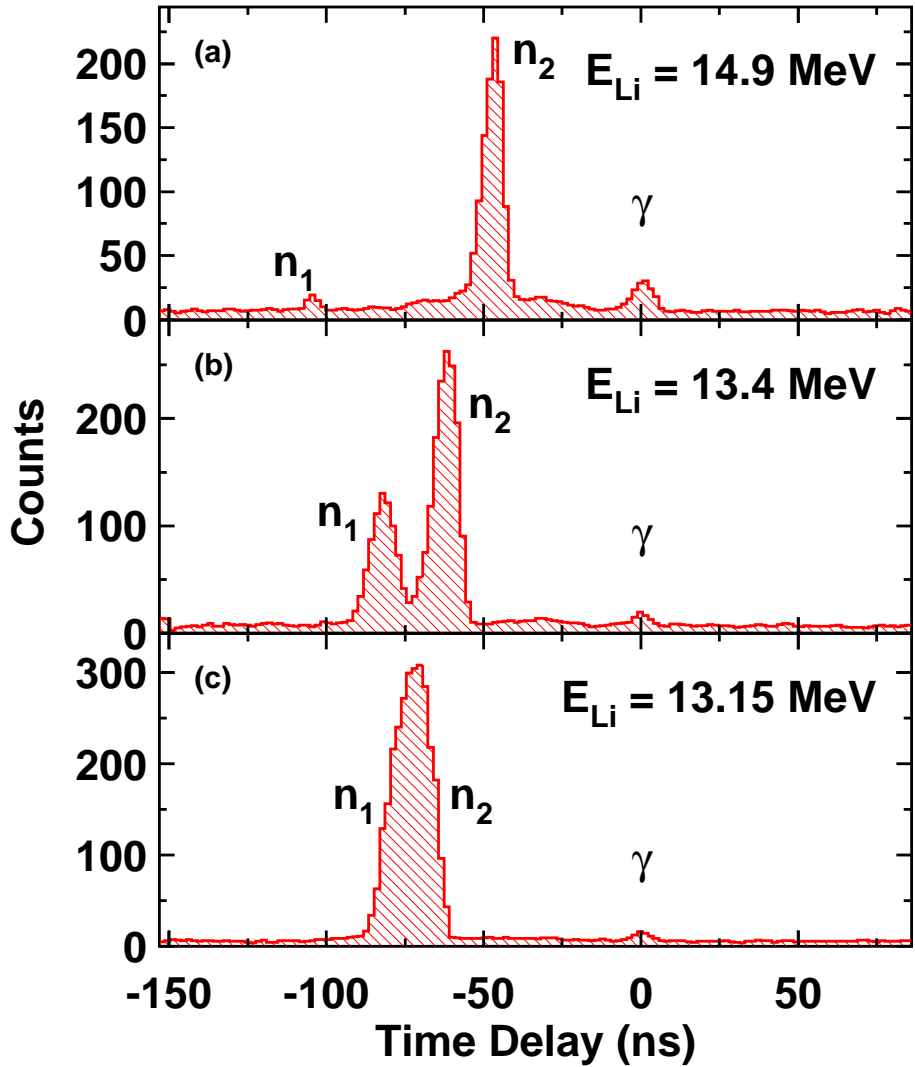
,height=17cm,bbllx=4cm,bblly=5cm,bburx=24cm,bbury=25cm}
\caption[]{
        Time of flight spectra between the arrival of the $^{7}$Li beam burst
        on the hydrogenated target and the detection of a particle in the
        NE213 neutron detector at $\Theta=0^o$ for
        $^{7}$Li beam energies equal to (a) 13.15 (b) 13.4 and (c) 14.9 MeV.}
\label{fig-lizero}
\end{figure}

As shown in Fig.~\ref{fig-lizero},
this background is found to be less important
with the $^{7}$Li beams, which were therefore
selected for the following experiments.
The selected $^7$Li beam energies for the
NaI and Ge measurements are 13.7 and 14 MeV,
corresponding to $n_2$ neutron kinetic energies at
$\Theta$=0$^o$ of 2.15 and 2.32 MeV and
maximum emission angles  $\Theta_{max}$ of
11.5$^o$ and 13.4$^o$, respectively.

The neutron beam energy resolution is dominated by
the energy losses in the target (150 keV over the entire
target thickness).
The contributions from the Li beam properties themselves
and straggling are 14 and 4 keV, respectively.
The resulting uncertainty on $E_n$ is 59 keV.
The calculated neutron flux~\cite{bib-dave}
is approximately of $10^5$ n/s/str for a $^7$Li
beam intensity of 4~nA.

\subsection{Neutron Detector Array}

In the quenching measurements, the NE213 detector at
$\Theta=0^o$ is removed, and the detector under
study is installed 60~cm away from the polyethylene target.
The scattered neutrons are detected in an array of 48 NE213
scintillators located at a distance of 100 cm from the
central detector and distributed in four rings corresponding to
scattering angles of 45$^{o}$, 90$^{o}$, 120$^{o}$ and 165$^{o}$.
Thus, in one experiment the quenching factor is measured at
four different values of $\theta$, and consequently at four
different values of $E_R$.
The recoil energy values probed in the NaI and Ge experiments
are listed in Tables~\ref{tab-nai} and~\ref{tab-ge}.

\begin{figure}[tbp]
\epsfig{file=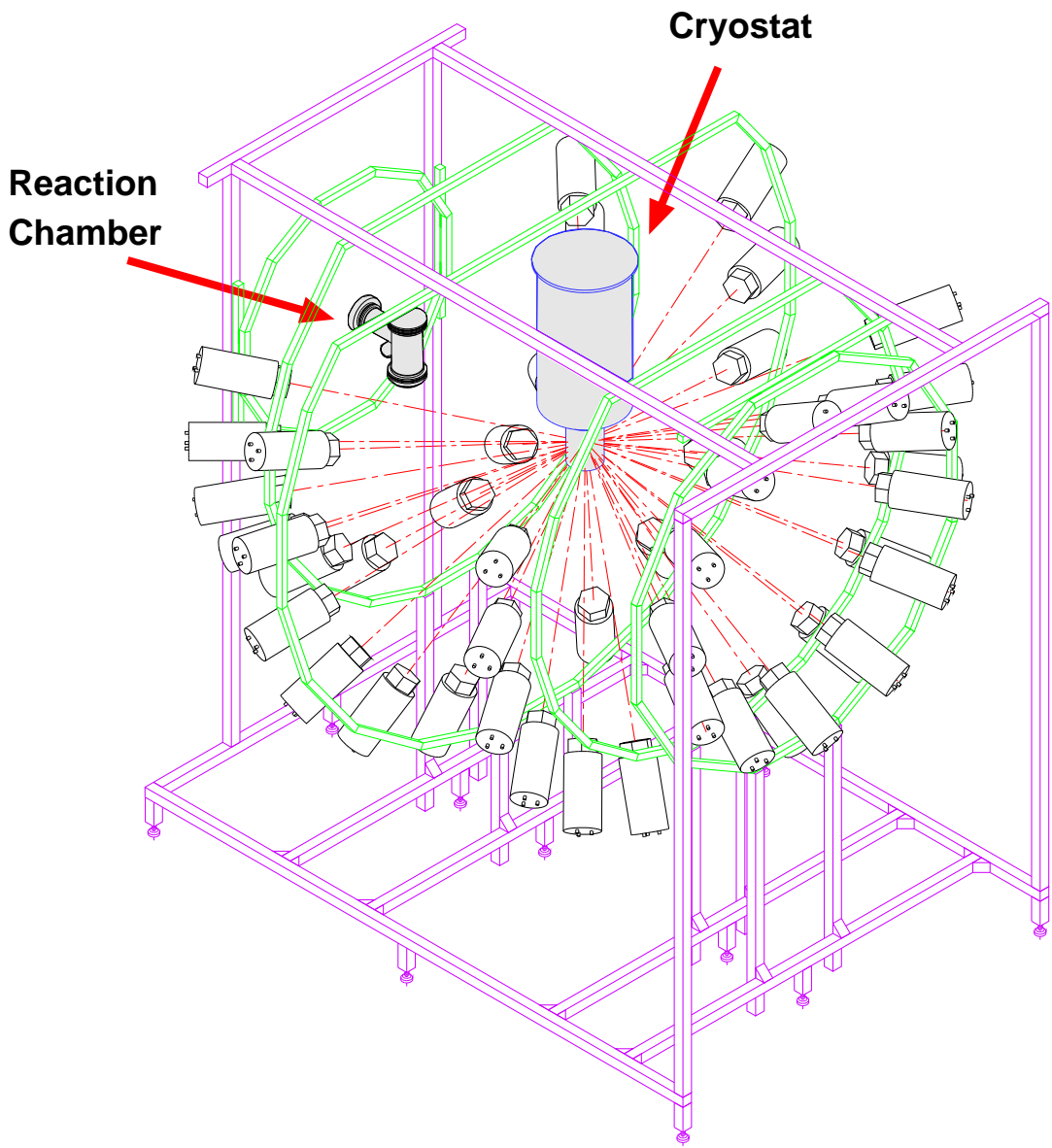
,height=17cm,bbllx=4cm,bblly=4cm,bburx=24cm,bbury=24cm}
\caption[]{
        Experimental setup
        with the four  rings (45$^o$, 90$^o$, 120$^o$
        and 165$^o$) of neutron detectors and its
        support structure.
        The cryostat and the reaction chamber are also shown.}
\label{fig-setup}
\end{figure}

The NE213 hexagonal cells have an external diameter of 10 cm and a
thickness of 4 cm.
A photomultiplier is optically coupled to each cell
via a light guide.
The support structure of the 48 neutron detectors
is shown in Fig.~\ref{fig-setup}.
It is made of dural and steel,
thin enough to minimize additional neutron scattering.
The detector positioning is performed in the following way.
First, a laser beam is propagated along the accelerator beam axis,
simulating the neutron beam at $\theta=0^o$.
A rotating mirror is then set at the center of the array.
The mirror is then tilted to the
desired scattering angle and is rotated around the beam axis.
The alignment of each NE213 detector is then adjusted so that the
center of its front face is illuminated by the reflected
laser beam.

\subsection{Electronics}

The background due to $\gamma$ interactions in the NE213
cells is reduced by using the neutron-$\gamma$
discrimination capabilities of these detectors.
The signal coming from each detector is sent to two charge digitizers,
integrating the current during T$_1$=40~ns and T$_2$= 400~ns,
respectively.
The ratio of the corresponding integrated charges $q_1$ and $q_2$
provides a good $\gamma$/neutron discrimination.

The energy signal $E_{ee}$ from the center detector
is obtained with standard spectroscopy amplifiers
in the cases of scintillation in NaI and ionization in Ge.
The much slower heat signals of the Ge cryogenic detector
are recorded with a wave form digitizer.

For each event, the following parameters are stored:
the $q_1$ and $q_2$ values for all NE213 cells,
as well as their time-of-flight relative to the beam burst,
and the energy and time-of-flight measurements
of the Ge or NaI(Tl) detector under study.
As it will be seen, the time delay between the beam
burst and the hit in the neutron detector is
a powerful tool to identify true neutron
scattering events.
This delay $\Delta t_{1-3}$ is the sum of those incurred as
the neutron goes
{\em i)\/} from the polyethylene target to the center
detector ($\Delta t_{1-2}$) and 
{\em ii)\/} from that detector to a NE213 cell
($\Delta t_{2-3}$).
With a fast center detector such as NaI,
additional reduction of random coincidences
may be achieved by a further discrimination
on either $\Delta t_{1-2}$ or $\Delta t_{2-3}$  

An event is defined as a coincidence between the center
detector and at least one of the 48 neutron detectors.
Data read-out is triggered by coincidences
between any NE213 detector and the center detector
within a time window of 1 $\mu$s.
This width is chosen so as to accommodate the time
resolution on the lowest energy signals in the center
detectors.

The most delicate aspect of the trigger is to reconcile
the speed of the NE213 signal read-out with the
slowness of the coincidence decision.
The solution adopted here is to initiate the digitization
of the NE213 information as soon as any of them
pass a certain threshold, and to hold this information
until the coincidence decision with the slower Ge
detector is issued.
If no coincidence arrives, the NE213 electronics is reset.
Otherwise, the NE213 data is further held until
the end of the Ge heat and ionization read out,
250 ms later.

For the tests with the faster NaI detector, the width of the
coincidence with the neutron array can be reduced to 500 ns,
and the stand-alone NE213 pre-trigger is not necessary.

\subsection{Recoil Energy Resolution}

The accuracy of the determination of $E_R$ on an event-by-event
basis depends on the uncertainty on $E_n$ and on $\theta$.
In natural Ge, an additional 2.6\% rms fluctuation on $E_R$ comes 
from the difference between the isotope masses.

The energy loss of the $^7$Li ions in the hydrogenated target
leads to a dependence of the neutron kinetic energy on the
position of the reaction along the target length.
For 2.32 MeV neutrons, the energy spread on $E_n$ is $\pm$3\%,
leading to an equivalent spread on $E_R$.
The main cause of event-by-event fluctuations of $\theta$ is
the solid angle of the NE213 cells.
Their contributions to the r.m.s. width of the $E_R$ distribution
decreases from 2.5\% at 90$^o$ to 0.6\% at 165$^o$.
All these effects taken together contribute to
a 4 \% r.m.s. event-by-event fluctuation of $E_R$.
This event-by-event uncertainty is much smaller than the resolution
that can be achieved by a time-of-flight measurement over a distance
of 1~m.
It is also significantly smaller than typical resolutions for the
measurement of $E_{ee}$ in NaI(Tl) or Ge detectors.


\section{Monte Carlo Simulations}

The experimental precision on $Q$ arises from the
possibility to select quasi-monoenergetic neutrons and from
the simple kinematics of elastic collisions.
The determination of $E_R$ depends only on the kinematics
of the neutron-producing reaction (yielding $E_n$)
and the accurate positioning of the NE213 cells
(yielding $\theta$).

To preserve this simplicity, it is important to reduce as
much as possible the multiple scattering of neutrons
and all sources of beam-correlated
signals between the target detector and the NE213 cells.
To investigate this, the experiment for the measurement
of the heat and ionization
quenching factors in Ge has been simulated
using the Monte Carlo program GEANT~\cite{bib-geant}.
In these simulations neutrons are produced one by one,
thus assuming that the beam current
and bunch structure is such that the pile-up of
interactions coming from two different neutrons
is negligible.
In addition to the Ge detector, the simulation takes into
account interactions occuring in the material composing
48 NE213 detectors and photomultipliers,
the cryostat and the reaction chamber.
It also takes into account of the
thickness of the hydrogenated target and
follows the flight of the
neutron from its creation to its detection
in one of the NE213 cells.
A cut on the time-of-flight similar to the one later applied
to the data is also applied to the simulated data.

These simulations take into account the inefficiencies of
the NE213 cells due to the experimental energy thresholds
and their effect on the detection of the elasticaly and
inelasticaly scattered neutrons
(energies of 2.2 and 1.6 MeV, respectively: see below).

Predicted rates for elastic and inelastic
scattering events, including multiple-scattering events,
were derived from these simulations, assuming
an integrated flux of 70 neutrons impinging on
the Ge detector per second, corresponding to a $^7$Li
beam current of 5 nA.
The single rate in the center Ge detector is
approximately 10 neutrons per second.

Table~\ref{tab-rate} lists the Ge-NE213
coincident rates
for the different event types described in the
following.

{\bf Inelastic Neutron Scattering}

In addition to elastic scattering, neutrons can have
inelastic interaction with Ge nuclei.
The end products of such a collision are {\em i)\/}
a nuclear recoil with a kinetic energy inferior
to that associated with elastic scattering,
{\em ii)\/} a scattered neutron with a reduced
energy and thus an increased time-of-flight to
the NE213 cells, and {\em iii)\/} a $\gamma$ emission
which may deposit some of its energy in the detector.
Given the neutron energy, the most important $\gamma$'s
to be expected~\cite{bib-ninel} are the
596 keV transition in $^{74}$Ge,
835 keV in $^{72}$Ge and 563 keV in $^{76}$Ge.
The simulation predicts that 82\% of these $\gamma$
escape detection, leaving only the neutron time-of-flight
as a mean to discriminate these
events from elastic scatters.
The cross-sections for each of these states are
of the order of 1~b~\cite{bib-ninel}, compared with an
elastic cross-section of the order of 2~b~\cite{bib-nel}.
Given the isotopic abundances in Ge,
the 563 keV yield is only a fifth of the 596 keV one.

The recoil energies produced by these
inelastic processes are
approximately 6 keV inferior to the
values for elastic scattering.
The differential cross-section of the inelastic
process has a pronounced maximum for backward-scattered
neutrons, and it is only at these angles that
it can result in an appreciable background.
At 120$^o$, the coincident rate of inelastic events is
approximately half of the elastic rate (Table~\ref{tab-rate}).
These two types of events can be identified
by their $\sim$10~ns difference in arrival time in the
NE213 cells, and thus can be treated separately.

In the 18\% of cases where the $\gamma$ is detected,
most of the time it will leave only part of its
energy via Compton scattering.
The resulting distribution of energy deposited
in the Ge will be flat, starting at the
recoil energy and extending up to $\sim$500 keV.

{\bf Multiple Interaction of a Neutron in Ge}

The interaction length of 2.32 MeV neutrons in Ge
is approximately 6~cm.
Despite the small size of the Ge detector chosen for the
experiment (3.1 cm$^3$), incident neutrons have a 20\% probability
to interact twice or more in the crystal.
The corresponding coincident rates are approximately
half of the elastic rate (Table~\ref{tab-rate}).
It is not reduced by the cut on the time-of-flight
of the scattered neutron which is not significantly
altered by the successive interactions.
However the angle of the outgoing neutron is uncorrelated with
the energy deposited in the Ge detector.
Some rejection of this background can be thus achieved,
as single-scattering neutrons detected at a given angle $\theta$
should correspond to a single $E_{ee}$ value within experimental
resolution, while the corresponding distribution for multiple
scatters is a continuum extending approximately
from $E_{ee}$ to twice $E_{ee}$.

{\bf Collisions in the Surrounding Materials}

Neutrons can also have multiple interactions,
with at least one occurring in the Ge and the other(s)
in the materials surrounding the detector.
In the SICANE setup these additional interactions
occur mainly in the cryostat\footnote{The total thickness
of the copper cryogenic shields is 4mm.}
and the chamber containing
the hydrogenated target.
Once the time-of-flight requirements are applied,
the coincidence rate of such events is predicted to be
of the order of the elastic rate.
However, for half of these events, the energy
deposited in Ge is only a few keV and will
not pass the trigger requirement.
For the other half, the deposited energy
is evenly distributed between 0 and $E_{ee}$

In order to treat adequately the background of events
coming from
{\em i)\/} inelastic recoils followed by Compton-scattering,
{\em ii)\/} multiple neutron scattering in the Ge
itself and
{\em ii)\/} multiple scattering in the surrounding material,
a smooth background is included in the fits to the
observed $E_{ee}$ distributions yielding the average
value at a given angle and time-of-flight selection.

\section{Scintillation Quenching Factor of NaI(Tl)}

As a first test of the performance of the SICANE
facility, it was used for a measurement the scintillation
quenching factor of NaI(Tl).
Measurements with NaI(Tl) are considerably easier to perform
than measurements with Ge.
There is no need for a cryostat
and in addition the good timing resolution
provides an additional
test to remove random coincidences:
the neutron time-of-flight between the
hydrogenated target and the NaI(Tl).

\begin{figure}[tbp]
\epsfig{file=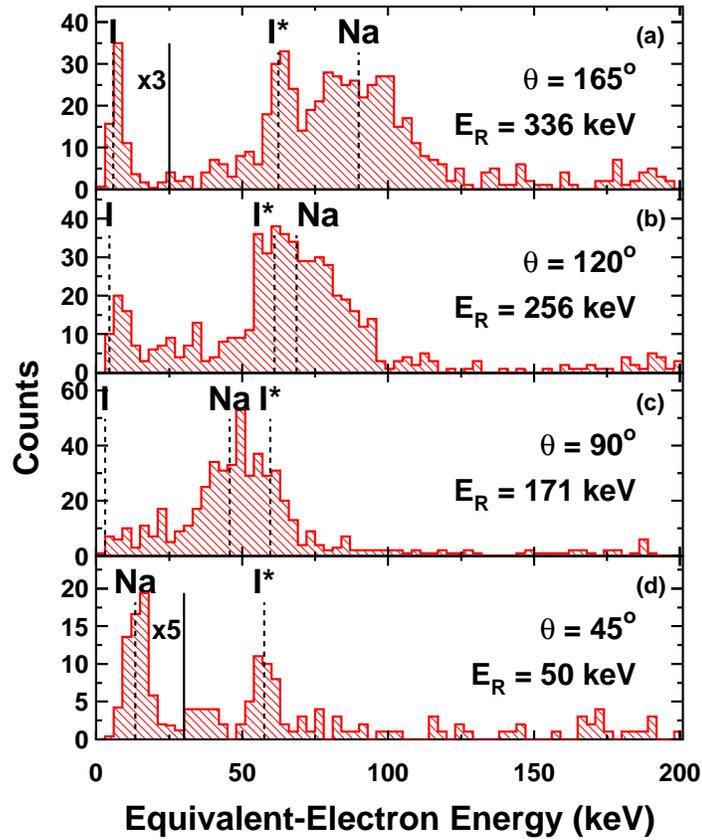,width=13cm}
\caption[]{
         Energy spectrum (in equivalent-electron keV) recorded
         in the NaI detector in coincidence with a neutron identified
         in a NE213 cell at $\theta$ = 165$^{o}$, 120$^{o}$, 90$^{o}$
         and 45$^{o}$ (a to d, respectively).
         The number of counts below 24 keV at 165$^o$ has been
         scaled down by a factor of 3 for clarity.
         The dashed lines show the expected peaks positions for
         elastic (Na,I) and inelastic (I$^*$) scatterings,
         assuming Q(Na)=0.27 and Q(I)=0.09.}
\label{fig-enai}
\end{figure}

In order to reduce multiple scattering,
the NaI(Tl) detector had a small size
(25~mm diameter, 25~mm thickness).
Its energy response $E_{ee}$ for photons was calibrated
in the energy range from 60 to 511 keV
using $^{57}$Co, $^{241}$Am and $^{22}$Na radioactive sources.
With incident neutron energy of 2.1 MeV
only sodium recoils are observed
at all four scattering angles of
$45^{o}$, $90^{o}$, $120^{o}$ and $165^{o}$,

To simulate the experimental conditions for the Ge experiment,
the total count rate in the NaI(Tl) detector was limited to 35~Hz,
yielding a coincidence count rate of 0.33~Hz.
Figure ~\ref{fig-enai} shows recoil energy spectra recorded at
different scattering angles after the selections
on the neutron identification parameter in the NE213 cells
and on the total neutron time-of-flight between the
polyethylene target and the NE213.
The largest peak observed in each spectrum corresponds to
the elastic scattering of neutrons on Na nuclei.
The peak around 60 keV, clearly seen at $\theta$=45$^o$,
is due to inelastic scattering via
the first excited state of $^{127}$I.
The structure below 10 keV evolving at the different angles could be due
to neutron elastic scatterings on $^{127}$I, producing a peak
expected at approximately 6 keV at $\theta$=165$^o$,
and at even lower energies at other angles.
An additional cut on the time-of-flight
between the NaI(Tl) and NE213 detectors does not remove
a significant number of events.
This indicates that the neutron identification and
the total time-of-flight are sufficient to select
a clean sample of neutron scattering events.

\begin{figure}[tbp]
\epsfig{file=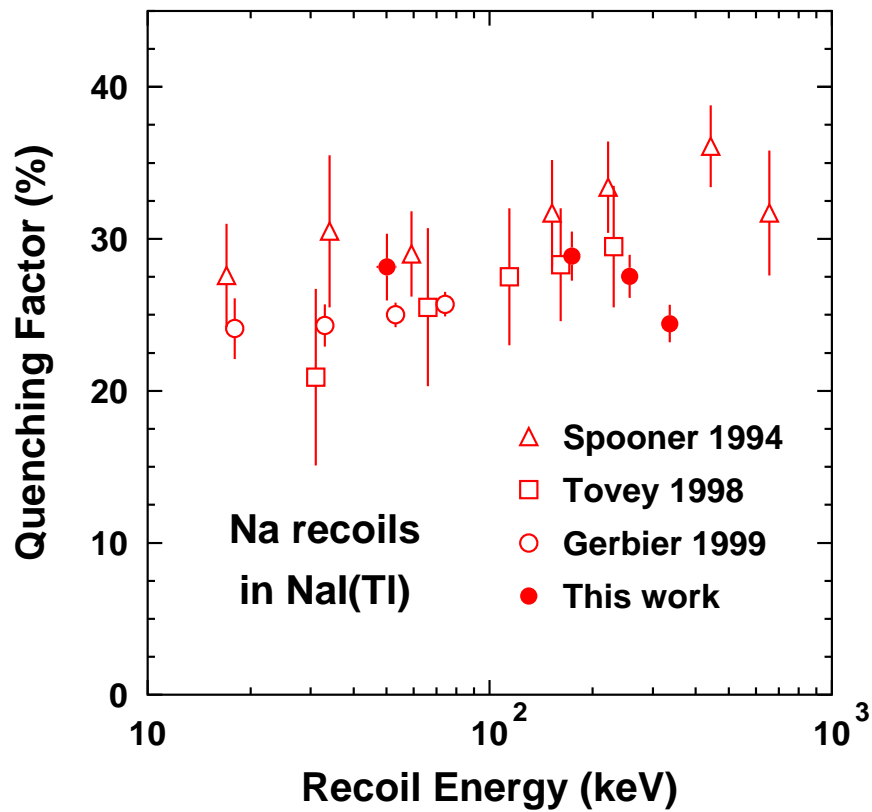,width=13cm}
\caption[]{
        Scintillation quenching factor for Na recoils
        in NaI(Tl) as a function of the recoil energy.
        The experimental data are from
        Refs.~\protect\cite{bib-nai}
        and from this work.
}
\label{fig-naiq}
\end{figure}

The $E_R$ values for Na recoils in elastic scattering
for the different scattering angle values and
the corresponding quenching factors values are listed
in table~\ref{tab-nai}.
As shown in Fig.~\ref{fig-naiq}, the measured quenching factors
are in good agreement
with previous results~\cite{bib-nai}.
Although the present measurement is based on a one-day test
experiment, the statistical errors are already below
the systematic uncertainties.
At low recoil energy, the dominant uncertainty is the recoil
energy due to the finite solid angle of the NE213 cells.
At high recoil energy, the main uncertainty is the
$\gamma$ energy calibration.

\section{Ionization and Heat Quenching Factors of a
         Germanium Bolometer}

\subsection{Detector principle}

   The EDELWEISS collaboration~\cite{bib-edw2000} searches for WIMP
Dark Matter using heat-and-ionization cryogenic germanium detectors, 
also called bolometers.
   The energy deposited in the detector produces two signals.
First, the electron-hole pairs created by ionization are collected
on electrodes implanted or vaporized on the surface of the crystal.
Secondly, a thermistance measures the few $\mu$K heating of the
crystal due to 
{\em i)\/} the thermalization of the electron or nuclear recoil
from the primary interaction, and
{\em ii)\/} the Joule energy associated with the migration
of the electron-holes due to the applied electric field.
Whether the initial recoil is electronic or nuclear,
the subsequent Joule heating (or Luke-Neganov
effect~\cite{bib-luke}) depends only on the
strength ionization signal, and thus is not affected by
the quenching factor.
Correcting for this effect in the presence
of a bias $V_o$ on the electrodes, the quenching factors
associated with the ionization and recoil energy
measurements ($Q$ and $Q'$, respectively) are
\begin{eqnarray}
Q  & = & \frac {E_I} {E_R} \\
Q' & = & ( 1 + \frac{V_o}{V_{pair}} ) \frac{E_H}{E_R}
             - Q\frac{V_o}{V_{pair}}
\end{eqnarray}
where  $V_{pair}$ = 2.96~V is the electron-hole pair creation
potential in Ge.
$E_I$ and $E_H$ are respectively the electron-equivalent
ionization and heat signal amplitudes,
i.e. as calibrated used $\gamma$-ray sources.
$E_R$ is the recoil energy as given by Eq.~\ref{eqn-ern}.

\subsection{Cryostat}

For the measurement of its quenching factor, the cryogenic
Ge detector is placed inside a dilution refrigerator.
A base temperature of 25 mK was achieved, however
it was regulated at 35 mK in order to optimize the
properties of the heat signal for this measurement.
The thicknesses of the different screens of the cryostat
were chosen to minimize neutron scattering.
Special care was taken to avoid mechanical vibrations.
In order to isolate the cryostat from the beam line components
and the multidetector array elements,
it is maintained at beam height (1.8~m) with a wooden
structure mounted on pneumatic insulators.
A metallic platform above the cryostat facilitates
liquid nitrogen and helium fills and access to
the bolometer electronics.
The center of the detector is 60 cm away from the
polyethylene target.

\subsection{Detector}

The bolometer used in this work is a cylindrical crystal
with a 10~mm height and a 20~mm diameter.
In this prototype detector~\cite{bib-stef} the heat sensor is a NbSi
thin film used as an Anderson insulator.
The operating potential $V_o$ is 6.26 V.
To facilitate the detector monitoring, a weak collimated
$^{57}$Co source is mounted close to it.

Keeping the data rate as high as possible is desirable
in this statistics-limited measurement.
In this respect, an advantage of NbSi thin film sensors relative
to the NTD sensors normally used by the EDELWEISS collaboration
(see e.g. Ref.~\cite{bib-edw2000}) is the relatively
fast risetime and decay of the heat signal.
These time characteristics depend on the temperature
of the detector.
At 35~mK, the risetime is approximately 1~ms and
the decay of the signal has a fast ($\sim$4~ms)
and a slow ($\sim$40~ms) components.
The fast component is due to ballistic phonons
and its amplitude depends not only on the deposited
energy but also on the interaction point relative to
the sensor.
The slow component corresponds to thermal phonons and depends
only on the deposited energy.
It limits the acceptable interaction rate in the detector to
approximately 15~Hz.

The electron-equivalent energy resolutions on ionization and heat
signals in this experiment are 8 and 41 keV FWHM at 122 keV,
respectively.

\subsection{Ionization quenching factor of Ge}

The ionization measurement has a trigger
threshold of 17 keV.
With a quenching value of approximately 30\%, the expected
ionization signal associated with neutron scattering in
the 90$^o$ ring is only 18 keV.
Consequently, the NE213 detectors were
grouped on the 120$^o$ and 165$^o$
rings, corresponding to recoil energies of 93
and 123 keV, respectively.

\begin{figure}[tbp]
\epsfig{file=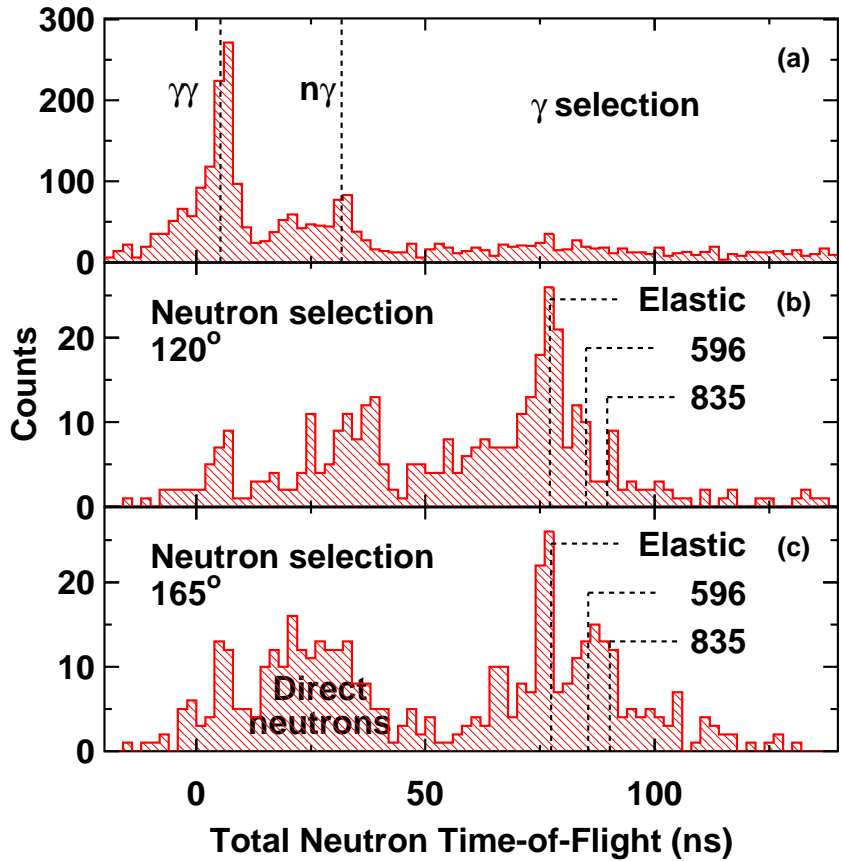,width=13cm}
\caption[]{
         Distribution of time-of-flights relative to
         the beam bursts, observed in the
         neutron detectors at 120$^o$ and 165$^o$
         when selecting (a) events identified as
         $\gamma$, (b)  events identified as neutrons
         at 120$^o$ and (c) at 165$^o$.
         The time intervals expected for
         ($\gamma,\gamma'$), (n,$\gamma$), (n,n) or
         (n,n') scattering in the Ge detector
         are indicated as dashed lines.
         Direct neutrons are those going directly from the
         hydrogenated target to the neutron detector
         without passing through the Ge detector.}
\label{fig-getof}
\end{figure}

Fig.~\ref{fig-getof} shows
the time-of-flight spectra between the polyethylene target
and the neutron detectors.
In Fig.~\ref{fig-getof}a, only those events rejected
by the n/$\gamma$ discrimination variable
of the NE213 detectors are displayed.
The peaks associated with photons produced either from
$\gamma$ or neutrons interacting in the Ge detector
(labeled $\gamma\gamma$ and $n\gamma$, respectively)
are clearly seen.
Accepted events in the 120$^o$ and 165$^o$ rings are shown
in Fig.~\ref{fig-getof}b and c, respectively.
The $\gamma\gamma$ and $n\gamma$ peak are strongly suppressed.
Neutrons associated with an elastic interaction in
the Ge detector are expected at time-of-flights of 77.1 and
77.4 ns, for the two respective rings.
The broad structure from 20 to 40 ns corresponds to neutrons
created in the polyethylene target hitting directly a NE213
cell without prior interactions.
This implies that the energy deposit in the Ge detector
is not related to the neutron in the NE213 cell,
and could be related to another particle emitted
by another interaction in the polyethylene target
in the same beam bunch.

The most important inelastic states expected to
be excited are the 563 keV states in $^{76}$Ge,
596 keV in $^{74}$Ge, and
835 keV in $^{72}$Ge.
The corresponding time-of-flights are 84.5, 85.1 and
89.7 ns at 120$^o$, respectively, and approximately 0.5 ns
later at 165$^o$.
An excess of counts is indeed observed at these time-of-flights
in Figs.~\ref{fig-getof}b and c,
especially at 165$^o$ where the simulations
predict the largest inelastic yields.

\begin{figure}[tbp]
\epsfig{file=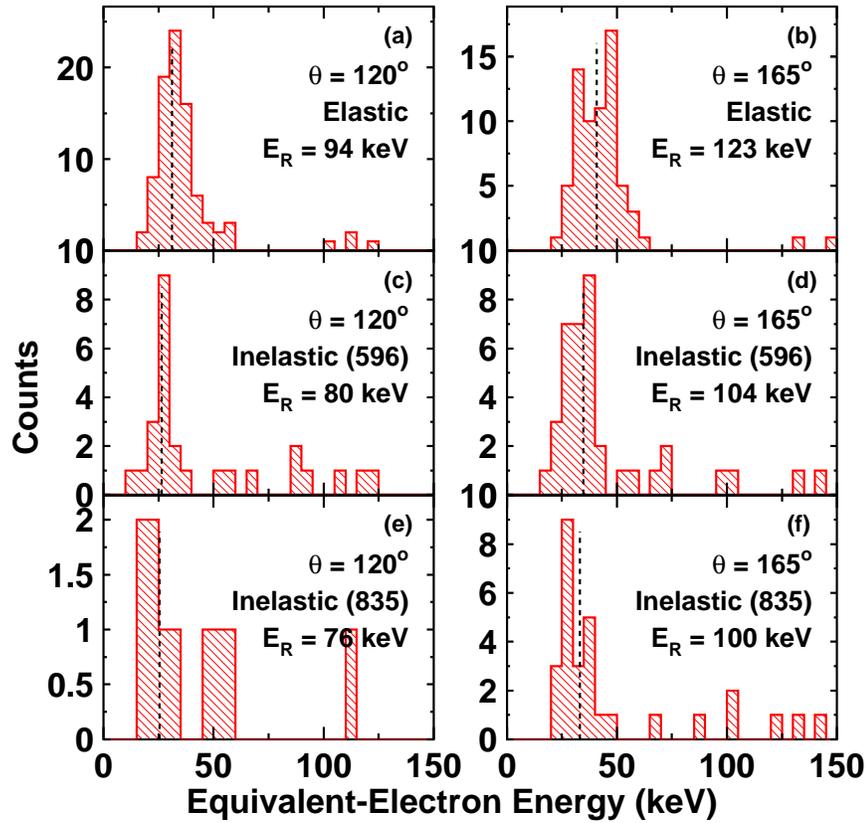,width=13cm}
\caption[]{
         Ionization energy spectra
         (in equivalent-electron keV) recorded
         in the Ge detector in coincidence with a neutron identified
         in a NE213 cell at $\theta$ = 120$^{o}$ (a,c,e at left)
         and 165$^{o}$ (b,d,f at right).
         The elastic selection (a and b) corresponds to
         total neutron time-of-flights of 70 to 80 ns,
         the ``inelastic (596)'' selection, to 80-87 ns (c and d), and
         the ``inelastic (835)'' selection, to 87-92 ns (e and f).
         }
\label{fig-geion}
\end{figure}

To clearly assess whether these events are indeed
due to inelastic scattering,
the events are categorized according
to 3 intervals of time-of-flight,
corresponding to 3 scattering processes:
70 to 80 ns (elastic scattering),
80 to 87 ns (inelastic scattering with E$_{\gamma}$=563
             or 596 keV), and
87 to 92 ns (inelastic scattering with E$_{\gamma}$=835 keV).
The associated ionization spectra for these
three categories at the two angles are shown
in Fig.~\ref{fig-geion}.
A peak of events at low ionization energy is observed
in each category.
Its mean energy is largest in the elastic selection,
and decreases with increased $E_{\gamma}$.
The peak locations are consistent with those
calculated (dashed lines) with the expected
recoil energy for each event category, assuming a common
quenching value of 33.3\%.
The fact that a flat background is more present
in Figs.~\ref{fig-geion}c to f than in
Figs.~\ref{fig-geion}a and b
provides further support for the identification
of inelastic scattering, as this behavior is
expected as emitted photon may deposit some of its energy
in the Ge detector.
However, the relative amount of smooth background
may also be partly due to the difference in the
signal-to-background for each event category
(as seen in Fig.~\ref{fig-getof}).

\begin{figure}[tbp]
\epsfig{file=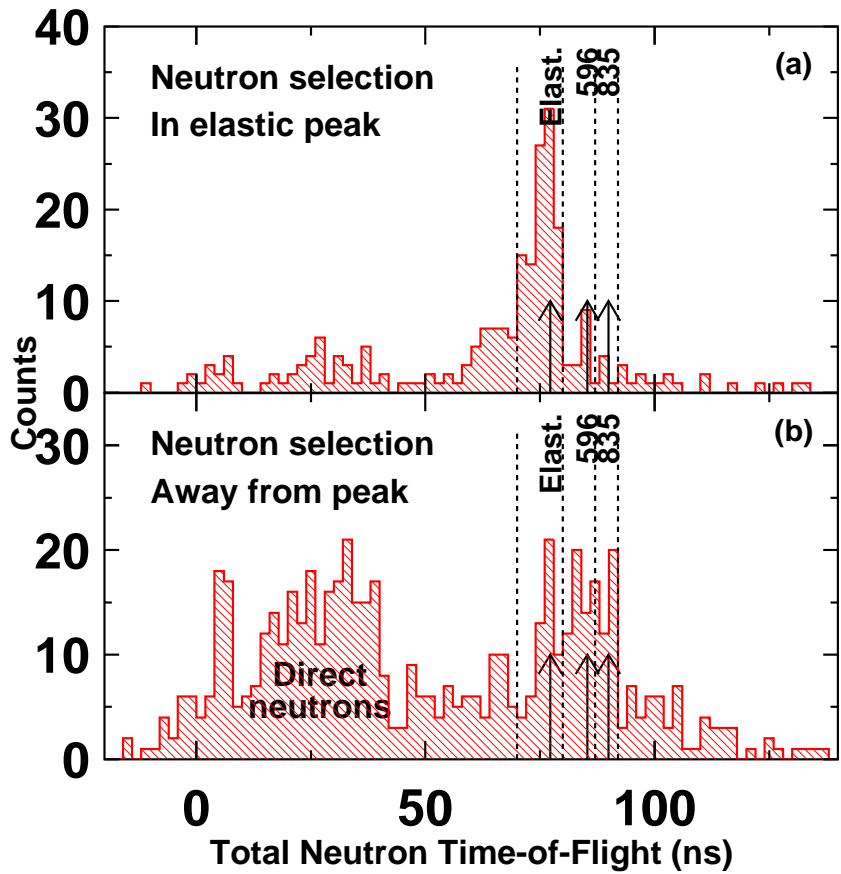,width=13cm}
\caption[]{
         Distribution of time-of-flights relative to
         the beam bursts, observed in the
         neutron detectors at 120$^o$ and 165$^o$
         for events identified as neutrons, with
         (a) the cut on the ionization signal
         in Ge that favors elastic events
         (see text) and (b) the cut favoring
         inelastic events.
         The expected time-of-flights of
         elastic and inelastic events are marked
         with arrows.
         The dashed lines represent the three
         time-of-flight intervals used in the
         previous figure.}
\label{fig-geinel}
\end{figure}

A final check of the assignment of the events
of Figs.~\ref{fig-geion}c to f to inelastic collisions
is to look back at the time-of-flight spectra of
the neutron-selected events, but this time
separating the events in two samples,
according to the recorded ionization signal.
Fig.~\ref{fig-geinel}a shows the time-of-flight spectra
of events with ionization energies clearly associated
with the peak observed in Figs.~\ref{fig-geion}a and b,
that is, between 25 and 50 keV at 120$^o$ and
40 to 65 keV at 165$^o$.
Fig.~\ref{fig-geinel}b shows the time-of-flight spectra
of events outside this ionization energy intervals.
As expected, the elastic scattering peak is well suppressed
in the second sample.

These tests clearly demonstrate the presence of inelastic
scattering events, and indicate that the time-of-flight
selection is adequate for identifying them, given
the present statistics.
This identification is important since the
difference in recoil energy between elastic and
inelastic scattering events is about 20\%.
A improper identification of the elastic sample 
leading to a contamination of inelastic events
could have lead to a significant bias in the measurement
of the quenching factor.
In our case, assigning the elastic value for the
recoil energy to all events in the 68 to 92 ns range
would have reduced the average Q value measurement by 5\%.

It should be noticed that, with sufficient statistics,
inelastic events increase the number
of recoil energies covered by the measurements,
going down to lower values (see Fig.~\ref{fig-geion}).

\begin{figure}[tbp]
\epsfig{file=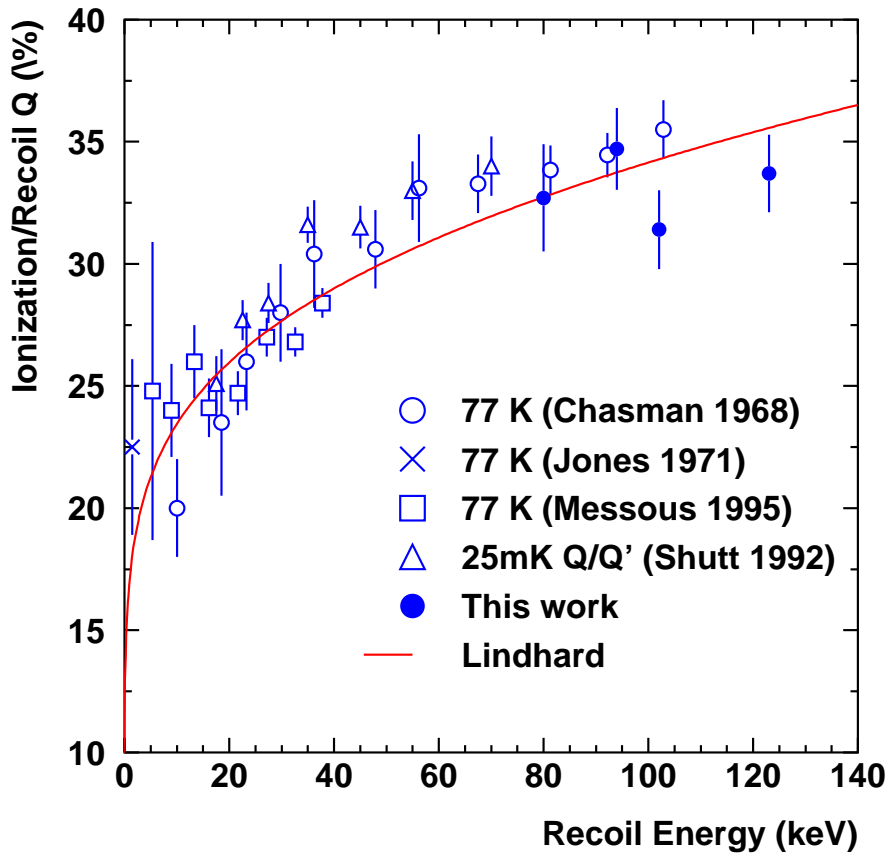,width=13cm}
\caption[]{
        Ionization quenching factor measured for
        Ge recoils in Ge(Li) detectors at liquid nitrogen
        temperature~\protect\cite{bib-chasman,bib-jones,bib-messous},
        and Ge recoils in hyperpure Ge at 35 mK
        (this work).
        The $Q/Q'$ measurement of Ref.~\protect\cite{bib-shutt}
        is also displayed, assuming a heat quenching $Q'$ of
        100\%.
        The thick line represents a calculation based
        on the Lindhard theory~\protect\cite{bib-lindh},
        as parametrized in Ref.~\protect\cite{bib-lewin}.}
\label{fig-qion}
\end{figure}

The quenching values extracted from Fig.~\ref{fig-geion}
data are listed in table~\ref{tab-ge} and shown on
Fig.~\ref{fig-qion}.
The statistical precision on the average $Q$ value
obtained in a one-day experiment is 2\%.
The results are consistent within statistical uncertainty
with published results~\cite{bib-chasman,bib-shutt,bib-messous}
and the Lindhard model calculations.
No significant systematic shifts are observed between the
measurements of Ge(Li) in liquid nitrogen
or our measurement of hyperpure Ge at cryogenic temperatures,
or between our direct measurement of $Q$
and the $Q/Q'$ measurement of Ref.~\cite{bib-shutt}
in the same temperature range.

\subsection{Heat quenching factor of Ge}

The analysis for the heat quenching follows closely that
of the ionization channel.
The neutron and time-of-flight selections are identical.

However, in addition to the significantly worse
resolution of the heat signal relative to ionization,
the slow components of the heat signals are strongly affected
by pile-up and microphonics.
In order to get a heat-to-ionization ratio of one for events
clearly associated with photons (ionization above 100 keV
and identification in the NE213 cell),
it is necessary to impose that the energy deduced from
the slow and fast components agree to within 75 keV.
This ensures the integrity of the adjusted pulse shape,
at the price of rejecting all surface events with anomalously
large fast components.
Clearly, the problem of pile-up is a strong limitation
for quenching measurements with slow cryogenic detectors.

Given the reduced statistics, a meaningful measurement is
only obtained for the total sample, dividing each measured
energy by the recoil energy corresponding to the given
angle and time-of-flight selection.
The resolution does not allow for the identification
of those inelastic events where the photon was partly detected.
In order to be able to use inelastic events despite this
problem, events with {\em ionization\/} quenching values
below 20\% and above 45\% are rejected.
 
\begin{figure}[tbp]
\epsfig{file=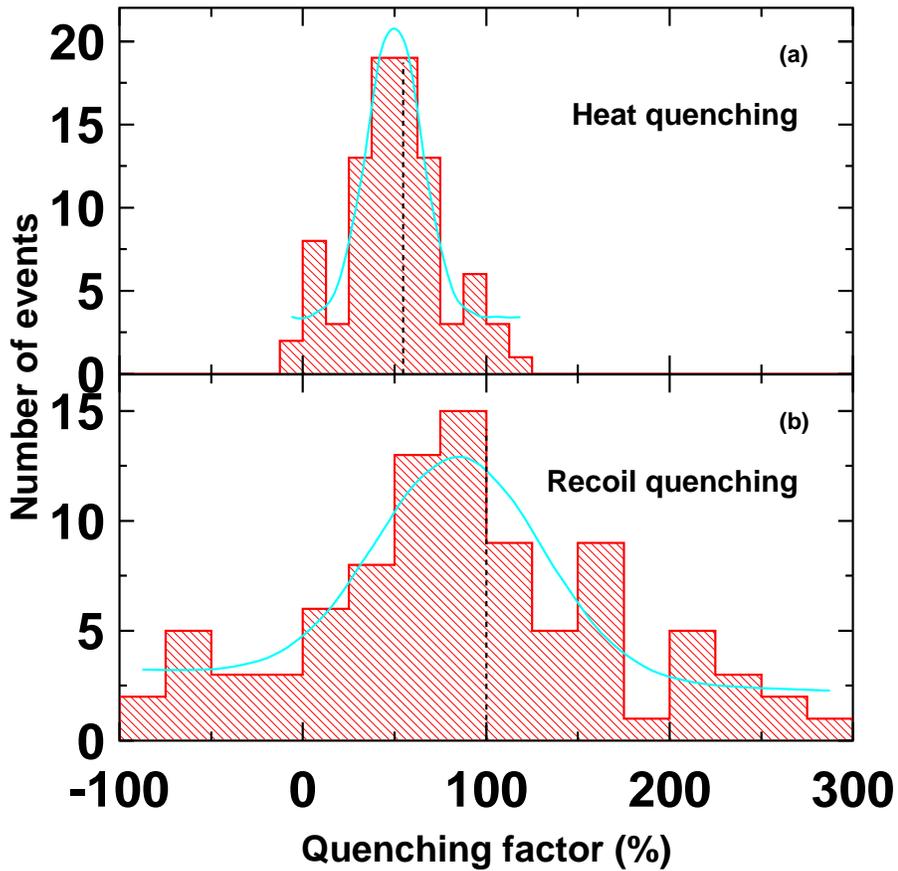,width=13cm}
\caption[]{
           Quenching factor for (a) the heat signal
           and (b) the recoil energy deduced from
           the difference between the heat and ionization
           signals.
           The full lines correspond to fits to the data
           yielding the values discussed in the text.
           The dashed lines represent the expected values
           assuming $Q$ = 33.3\% and $Q'$ = 100\%.
           }
\label{fig-gechal}
\end{figure}

The resulting distribution of heat quenching values is shown in
Fig.~\ref{fig-gechal}a.
A Gaussian fit to the distribution gives an average
quenching of 50 $\pm$ 3 \%.
Subtracting the Luke-Neganov effect
corresponding to the applied electric field of 6.24 V
using our measured average quenching value of 33.3\%,
we obtain (Fig.~\ref{fig-gechal}b) a quenching factor
for the nuclear recoil part of the signal
of 87 $\pm$ 10\%.
The quoted error is statistical only;
the systematic error,
mostly due to calibration uncertainties,
is of the same order.
Within these large errors,
the result is consistent with the expected value,
which should be very close to 100\%~\cite{bib-shutt,bib-aless}.

This one-day experiment clearly illustrates
what can be achieved with this technique.
The present statistical error would scale with
the improvement of resolution in the heat channel.
With sufficient energy resolution and statistics,
a measurement at the percent level may be possible,
if systematic effects such as inelastic scattering
are kept under control.

\section{Conclusion}

The commissioning tests of the SICANE array and its setup
confirm its high relevance for the calibration of the response
of cryogenic detectors to nuclear recoils.

Using inverse reactions to produce a highly collimated
neutron beam, no massive shielding of the neutron
detectors is necessary, giving more freedom in their
spatial setup.
Competitive results have been obtained in a short running
time (less than a day) for the quenching of scintillation
in NaI(Tl) and of ionization in Ge.
The array also makes possible a deeper investigation of the
effect of inelastic scattering on the measurement.

The powerful diagnostic tools provided by the array
(simultaneous measurements at different angles,
neutron identification and time-of-flight measurements)
are ideally suited for the study of cryogenic
detectors.
For such measurements, important factors that
require proper attention are:
{\em i)\/} the slow response of the detector, and how it compares
with the single rate in the cryogenic detector and
the rate in coincidence with the neutron array;
{\em ii)\/} the rate in the neutron detector array due to neutron
scattering in the material in the vicinity of the detector
and {\em iii)\/} the ability to resolve inelastic and
elastic events, either using the energy or the time-of-flight
measurements.

In addition, for the first time, it was directly verified that
the ionization quenching factor of Ge at 35 mK is very
close to that at liquid nitrogen temperature.
This is consistent with the fact that 
the CDMS and EDELWEISS collaborations~\cite{bib-shutt,bib-benoit}
obtain $Q/Q'$ ratios compatible with the $Q$ values measured
at liquid nitrogen temperature~\cite{bib-chasman,bib-messous}
under the assumption that $Q'$=1.
Conversely, this last hypothesis is verified for the first time
in the present work, albeit at a 15\% level.

\section{Acknowledgment}
We thank the technical staff of the Tandem accelerator of Orsay,
in particular
D. Gardes,
B. Waast and
J.M. Curaudeau
for their invaluable help.
This work has been partially funded by the EEC-Network program
under contract ERBFMRXCT980167.

\clearpage

\begin{table}[tbp]
\centerline
{\begin{tabular} {|c|c|c|}
\hline $\theta$  & $E_R$(Na)     & $Q$ \\
                 & (keV)         &  (\%) \\
\hline $45^{o}$  &   50 $\pm$  3 &  28.1 $\pm$ 2.8  \\
\hline $90^{o}$  &  171 $\pm$  6 &  29.4 $\pm$ 2.9  \\
\hline $120^{o}$ &  256 $\pm$  8 &  27.5 $\pm$ 2.8  \\
\hline $165^{o}$ &  336 $\pm$ 10 &  25.4 $\pm$ 1.2  \\
\hline
\end{tabular}}
\caption[]{
    Quenching values measured for Na recoils induced by the elastic
    scattering of 2.1 MeV neutrons for the different scattering
    angle values.
}
\label{tab-nai}
\end{table}

\begin{table}[tbp]
\centerline
{\begin{tabular} {|l|c|c|c|}
\hline
                  &  \multicolumn{3}{c|}{ Coincidence rate}         \\
       Process    &   \multicolumn{3}{c|}{(/hour)}            \\\cline{2-4}
                  &  165$^o$    & 120$^o$ & 90$^o$             \\
\hline 
  Elastic         & 2.0         &  1.6   &  0.2              \\
  Inelastic       & 1.7         &  0.6   &  0.1              \\
\hline
  Multiple scattering inside Ge  & 1.0        &  0.6   &  0.1      \\
  Multiple scattering outside Ge & 1.9        &  2.8   &  0.3      \\
\hline
\end{tabular}}
\caption[]{
    Simulated rates of Ge-NE213 coincidences after neutron
identification and requiring that the neutron time-of-flight
from the hydrogenated target to the NE213 counter is between
70 and 85 ns.
The assumed incident yield on the target is 70 neutron/s.
}
\label{tab-rate}
\end{table}

\begin{table}[tbp]
\centerline
{\begin{tabular} {|c|l|c|c|}
\hline $\theta$  & Process     & $E_R$(Ge)     & $Q$  (\%)          \\
                 &             & (keV)         & (Stat. error only) \\
\hline $120^{o}$ & Elastic     &   94 $\pm$ 3 &  34.7 $\pm$ 0.8 \\
                 & Inel. (596) &   80 $\pm$ 2 &  32.8 $\pm$ 0.9 \\
                 & Inel. (835) &   76 $\pm$ 2 &  30.1 $\pm$ 4.0 \\
\hline $165^{o}$ & Elastic     &  123 $\pm$ 4 &  33.7 $\pm$ 0.7 \\ 
                 & Inel. (596) &  104 $\pm$ 3 &  31.4 $\pm$ 1.2 \\
                 & Inel. (835) &  100 $\pm$ 3 &  31.4 $\pm$ 1.4 \\
\hline           & Average     &              &  33.3 $\pm$ 0.5 \\
\hline
\end{tabular}}
\caption[]{
    Ionization quenching measured for Ge recoils induced by the elastic
    scattering of 2.32 MeV neutrons, for the different scattering
    angle values and scattering processes.
    The global systematic error on $Q$ due to calibration
    and beam energy uncertainties is 3\% (not included here).
}
\label{tab-ge}
\end{table}

\clearpage

\clearpage

\clearpage

\clearpage

\clearpage

\clearpage

\clearpage

\clearpage

\clearpage

\end{document}